\documentclass{aastex}
\usepackage{emulateapj5}

\slugcomment{To appear in ApJ}
\shortauthors{S. Molendi \& F. Pizzolato}
\shorttitle{Is the gas in cooling-flows multi-phase?}

\begin{document}
   
\title{Is the gas in cooling-flows multi-phase?}
\author{Silvano Molendi\altaffilmark{1} \& Fabio Pizzolato\altaffilmark{1,2}
}
\altaffiltext{1}{Istituto di Fisica Cosmica, CNR, via Bassini 15,
I-20133 Milano, Italy}

\altaffiltext{2}{Universit\`a degli Studi dell'Insubria, Polo di Como,
Dipartimento di Scienze Chimiche, Fisiche e Matematiche, Via Valleggio 11,
I-22100 Como, Italy}

\begin{abstract}

Employing XMM-Newton EPIC data we perform a detailed
comparison between different spectral models to test
whether the gas in cooling-flows is multi-phase or not.
Our findings all point in the same direction, 
namely that gas in cooling-flows does not show the wide distribution
of temperatures expected from standard multi-phase models.
This result has profound implications 
for cooling-flow models. Firstly, the large absorption column 
depths inferred by previous analysis of cooling-flow spectra 
are most likely an artifact following from the application 
of an incorrect spectral model to the data. Secondly, 
the mass deposition and mass flow
are likely to be much smaller than previously thought. 
Last, but perhaps not least, the term "cooling-flow"
cluster is probably no longer appropriate, as  it 
describes a phenomenon of smaller entity and impact than 
previously thought. 
We propose to substitute it with that of "cool-core" cluster. The 
latter definition is less ambitious than the first, as it 
reflects only an observational fact rather than an inferred 
physical property, the flow, but has the undeniable advantage 
of being firmer. 
 
\end{abstract}


\section {Introduction}

A large fraction of the clusters which 
have been imaged with X-ray telescopes show clear
evidence of a centrally  peaked  surface brightness 
profile. The gas properties derived by deprojecting the 
X-ray images of these systems imply central cooling 
timescales which are substantially smaller than the Hubble
time. In the absence of heating mechanisms, or of 
other forms of support, the cooling gas must flow inwards
and give rise to a phenomenon known as cooling-flow.

Further evidence in favor of cooler gas in the core
of clusters has come from X-ray spectra.
ASCA (e.g. Virgo/M87, Matsumoto et al.1996; Centaurus, 
Fukazawa et al. 1994) and 
BeppoSAX (e.g  Perseus/NGC1275, Molendi 1997;
Virgo/M87 Guainazzi \& Molendi 1999) spectra of cooling-flow 
clusters show evidence of a temperature decrement 
in their cores.
From spectral fitting  of ASCA data with multi-phase 
models (see further on for a description of multi-phase models) 
evidence of intrinsic absorption in the soft
energy band, from cold gas with typical column depths of 
a few 10$^{21}$ cm$^{-2}$, has been found (e.g. Allen 2000).

Early observations (e.g. Fabian, Nulsen and Canizares 1984) 
have shown that, although very peaked, the surface brightness
profile is not as peaked as it should be if all the cooling 
gas were to flow to the center. This implies that the mass 
deposition rate $\dot M$ is not the same at all radii but  
scales linearly with the radius, $ \dot M \propto r $, 
(e.g. Peres et al. 1998). 
The above result has been explained in the context of 
inhomogeneous models (Nulsen 1986, Thomas et al. 1987). 
In these models the gas is considered to be highly multi-phase, 
i.e. at each radius different phases, characterized by different 
temperatures and densities, coexist with one another. 
The different phases are kept in pressure equilibrium,
since the sonic timescale, on which 
pressure waves propagate, is shorter than the cooling 
timescale. 
The phases comove with typical velocities substantially 
smaller than the sound speed.
A chaotic magnetic field, which is supposed to be present in
the cores of cooling clusters, may help in threading the phases
together. The same B field is invoked to suppress heat conduction
which would otherwise rapidly bring the different phases to the 
same temperature. 
At temperatures of the order of 10$^6$ K the cooling timescale
becomes shorter than the sonic timescale and the gas blobs 
which reach this temperature fall out of pressure equilibrium
with the other phases. The cool blobs decouple from the flow
which continues to move inwards. 

Cooling-flows are frequent in clusters and must therefore 
be a persistent rather than an episodic phenomenon in the 
life of galaxy clusters. 
The amount of gas expected to cool from the X-ray temperatures
throughout the lifetime of a cluster is of the order 
$M=10^{12} M_{\odot} (\dot M /~100 M_{\odot}/yr)$ (Sarazin 1988). 
This mass, although smaller by orders of magnitude
than the total mass of a rich cluster, is by no means
small, indeed it is comparable to the mass of the cD galaxy 
at the center of cooling-flow clusters.
Searches longward of the X-ray band at optical (e.g. Donahue \& Stocke 1994) 
infrared (e.g. O'Dea et al. 1994) wavelengths and at 
21 cm (e.g. Dwarakanath, van Gorkom, \& Owen 1994) 
have systematically come up with mass estimates of cold gas which 
are orders of magnitude smaller than the mass which is estimated 
to be cooling from the X-ray band. This point, plus the lack of
direct evidence of the gas motion (X-ray 
instruments with spectral resolution orders of magnitude better
than what we now have are required),  have led some
to view cooling-flows suspiciously and to look for alternative
solutions. Many authors have suggested that the 
cooling might be balanced by a heating mechanism (e.g.
 Tucker \& Rosner 1983, Binney \& Tabor 1995, Ciotti \& Ostriker 2001). 
However no stable heating process yet devised is able
to counteract the effects of radiative cooling and account for
observed X-ray images (e.g. Canizares et al. 1993).

With the coming into operation of the latest
generation of X-ray observatories  XMM-Newton  and 
Chandra,  new light is being shed on cooling-flows. 
Observations of three clusters: A1835 (Peterson et al. 2001),
Sersic 159-03 (Kaastra et al. 2001) and A1795 
(Tamura  et al. 2001) with the RGS instrument on board 
XMM-Newton, have brought some unexpected results.
The RGS spectra of the above systems  all
show a remarkable lack of emission lines
expected from gas cooling below 1-2 keV. The most 
obvious interpretation is that gas is cooling down to
2-3 keV but not further. Moreover from the XMM-Newton 
EPIC observation of M87 (B\"ohringer et al. 2001) and 
the Chandra observation of Hydra A (David et al. 2001)
evidence has been found against multi-phase gas in the 
cooling-flows of these two clusters.

In this paper we make use of the unprecedented combination of
spectral and spatial resolution and high throughput of the
EPIC experiment on-board XMM-Newton to make a systematic 
comparison between different spectral models for the  
cooling-flows in  A1835, A1795 and M87/Virgo.
Our goal is to test if the gas is strongly multi-phase, i.e.
characterized by a broad range of temperatures extending down to
$\sim 0.1$ keV, or not.

\section {Observations and Data Preparation}

We use data from three  clusters  observed 
in the CAL/PV phase, namely  A1835, A1795 and 
M87/Virgo.
Details on the observations, as well as results from the 
analysis of these clusters, have already been 
published in the recent A\&A special issue. 
The A1835 results are reported in Peterson et al. (2001), 
the A1795 results in Tamura et al. (2001) and Arnaud et al. 
(2001), and the M87 results in B\"ohringer et al. (2001) 
and Belsole et al. (2001).  

In this  paper we do not make use of PN data.
From a comparative analysis of PN and MOS spectra
of 10 sources observed during the 
CAL/PV phase (Molendi 2001) we have found that the PN 
spectra give systematically softer
best fitting spectral models than the MOS
in  the 0.5-1.0 keV range.
If the  parameters of the model adopted to 
fit simultaneously the PN and MOS spectra are forced to be the same,
with the exception of the column density and the 
normalization, the best fitting PN column densities are 
systematically smaller than the MOS column densities.
For the 5 out of 10 objects in our sample where 
no excess absorption is expected, we have found
the best fitting column densities derived from the MOS 
to be always 
in good agreement with the line of sight galactic values, 
while the ones  estimated from PN data are always 
substantially smaller.
We conclude that, of the two instruments, the MOS has 
a more reliable calibration at soft energies.
Since the soft X-ray emission plays a central role 
in this  paper we have decided to limit our analysis 
to the MOS data.
 

We have obtained calibrated event files for the MOS1 and MOS2 
cameras for all three observations with SASv5.0. 
Data were manually screened to remove any remaining bright 
pixels or hot columns. Periods in which the background is 
increased by soft-proton flares have been excluded using an 
intensity filter; we rejected all events accumulated  when 
the count rate exceeds 15 cts/100s in the $[10-12]$ 
keV band.
We have accumulated spectra in concentric annular regions
centered on the emission peak with bounding radii:
0-0.5 arcmin, 0.5-1 arcmin and 1-2 arcmin
for A1835; 0-0.5 arcmin, 0.5-1 arcmin, 1-2 arcmin  
and 2-3 arcmin for A1795; 
0.5-1 arcmin, 1-2 arcmin, 2-3 arcmin,
3-4 arcmin, 4-5 arcmin, 5-7 arcmin and 7-9 arcmin
for M87. The 0-0.5 arcmin region of M87 has been excluded
as it is highly contaminated by the emission of the 
nucleus and the jet.
For all 3 clusters we have removed point sources.
For M87 we have also removed the substructures which
are clearly visible from the X-ray image (e.g. Belsole
2001). 
Spectra have been accumulated for MOS1 and MOS2 independently.
The Lockman hole observations have been used for the background. 
Background spectra have been accumulated 
from the same detector regions as the source spectra.

The vignetting correction has been applied to the spectra 
rather than to the effective areas, similarly
to what has been done by other authors who 
have analyzed EPIC data (Arnaud et al. 2001).

Spectral fits were performed in the 0.5-10 keV band for
A1835 and A1795 and in the 0.5-4.0 keV band for M87.
Data below 0.5 keV were excluded to avoid residual calibration 
problems in the MOS response matrices at soft energies.
Data above 4 keV were excluded from the analysis of M87
because above this energy the spectra show a substantial 
contamination from hotter gas emitted further out 
in the cluster, on the same line of sight. 
      
\section {Spectral Modeling}

All spectral fitting has been performed using version 
11.0.1 of the XSPEC package.
All spectral models include a multiplicative component
accounting for the line of sight galactic absorption,
which is fixed to the value derived from radio
maps. The adopted numbers  are $2.3\times 10^{20}$cm$^{-2}$,
$1.2\times 10^{20}$cm$^{-2}$ and $1.8\times 10^{20}$cm$^{-2}$
for A1835, A1795 and M87, respectively.
The values for A1835 and A1795 come from the maps of 
Dickey \& Lockman (1990) while the one for M87 is from 
a detailed study of the core of the Virgo cluster from
Lieu et al. (1996).
We note that the absorbing column depth towards
the field which has been used for background, 
i.e. the Lockman hole, is smaller,  $N_H \sim 6\times 10^{19}$ cm$^{-2}$, 
than that found on the line of sight of the 
3 clusters we are considering. 
In principle, this could lead to an underestimation of 
the background at soft energies and to a background 
subtracted source spectrum that is softer than the real one.
However, for  A1835, where the $N_H$ difference is largest,
it will lead to a background variation of less
than 15\% at 0.5 keV and of only a few percent
at 1 keV. The effect on the source spectra, which at energies
between 0.5 and 1.0 keV, are always at least 20 times more intense
than the background spectra,  will therefore be contained
to less than 0.75\% at all energies and for all spectra. 
Moreover, in the case of M87 and A1795, 
where the $N_H$ difference leads to a background variation of less
than  10\% at 0.5 keV, the effect on the source spectra
will be smaller than 0.5\% at all energies and for all spectra.

We have compared our data with three different 
spectral models.
Model A is a single temperature model (mekal model in XSPEC),
this model has 3 free parameters: the temperature $T$, the metal
abundance $Z$, which is expressed in solar units, and the normalization.
Model B includes a single temperature component 
plus an isobaric multi-phase component multiplied by an absorber
located at the source ( mekal + zwabs*mkcflow in XSPEC),
this model has 5 free parameters: the temperature $T$, the abundance 
$Z$ and the normalization of the single-phase component plus the 
normalization of the multi-phase component, which is expressed 
in units of the mass deposition rate, $\dot M$, and the column density 
of the intrinsic absorber, $N_H$. The other parameters of the multi-phase
component are not free: the minimum temperature, $T_{min}$, is fixed at a 
value of 100 eV, the maximum temperature, $T_{max}$, and the metal 
abundance, $Z_{mkcflow}$, are linked respectively to the temperature,
$T$, and the metal abundance, $Z$, of the single-phase component.
Model B  is  commonly adopted to fit spectra
from multi-phase cooling-flows (e.g. Allen 2000). This model was 
originally written to fit spectra from the
entire cooling-flow region. Since we are applying it to 
annular regions the normalization cannot be interpreted
as the mass deposition rate for the whole cooling-flow. However, the 
important point is that such a spectral component has the 
spectral shape expected form a multi-phase gas where the 
different phases are in pressure equilibrium with each other.
Model C includes a single temperature component 
plus an isobaric multi-phase component
 (mekal + mkcflow in XSPEC). 
As for model B this model has 5 free parameters:
the temperature $T$, the abundance 
$Z$ and the normalization of the single-phase component plus the 
minimum temperature, $T_{min}$, and the normalization of the 
multi-phase component, $\dot M$.
The other parameters of the multi-phase
component are not free: the maximum temperature, $T_{max}$ and the 
metal abundance, $Z_{mkcflow}$, are linked respectively to the 
temperature, $T$, and the metal abundance, $Z$, of the single-phase 
component.
This model is similar to the one adopted to fit the RGS data of
A1835 (Peterson et al. 2001). It may be regarded as a "fake multi-phase"
in the following sense.
From ASCA (e.g. Virgo/M87 Matsumoto et al.1996; Centaurus, 
Fukazawa et al. 1994) and 
BeppoSAX (e.g  Perseus/NGC1275 Molendi 1997; Virgo/M87 Guainazzi 
\& Molendi 1999) and more recently 
Chandra (e.g. Hydra A, David et al. 2001)  and XMM-Newton (e.g. 
A1835, Peterson  et al. 2001) observations we know that
clusters feature a temperature decrement within the cooling-flow 
region. 
Suppose now that the gas in cooling-flow clusters is 
not multi-phase but single-phase, than the spectra
accumulated from annuli where the temperature gradient is large
will appear as multi-phase not because the gas is truly multi-phase
but {\it because the observed spectrum comes from spatially distinct
regions characterized by different temperatures}.
Model C should give a better fit to the data than model B
if the gas is single-phase, because in this case the spectrum from 
a given annular region should be better described by a model
with a minimum and a maximum temperature, which describe 
the minimum and maximum single temperatures within
the given annular region, rather than from 
a model with a maximum temperature and a multi-phase component
with practically no minimum temperature.

In the case of M87 which, having a low temperature of 1-2 keV, 
features very prominent emission lines of various elements, 
we have allowed for independent fitting of individual metals abundances. 
This has been achieved by substituting the mekal and mkcflow components 
with vmekal and vmcflow respectively. Individual element
abundances for the vmcflow component, in models B and C,
have been linked to the same elements abundances in
the vmekal component.

On the basis of the above discussion it is relatively 
straightforward to make predictions as to what we should 
expect to find when comparing spectral fits obtained with 
models A, B and C.
If the medium is truly multi-phase, models B and C should give 
better fits than model A everywhere within the cooling-flow 
region. Model B should give better results than model C.
If the medium is single-phase then: for regions 
characterized by a narrow temperature distribution (i.e.
at relatively large radii),
models B and C should 
not provide substantial improvements with respect to
model A; for regions
characterized by a broad temperature distribution (i.e.
in the innermost radial bins) 
models B and C should provide substantial improvements
with respect to model A. Model C should give better
fits than model B.

\subsection{Results}


In the bottom panels of 
Figures 1, 2  and 3 we plot the temperature profiles for 
A1837, A1795 and M87 respectively, as obtained from model A and, 
for those bins where model C provides a significantly better fit than model 
A, the minimum and maximum temperatures obtained from model C.
In the top panels of the same figures we plot the  
$\Delta \chi^2$ between models A and B (filled circles) 
and models A and C (open circles) as a function of radius. 
The horizontal lines indicate 
the $\Delta \chi^2$ values for which the statistical improvement of 
the model B or C fit with respect to model A are significant at the 
99\% level according to the F-test. Whenever a datapoint 
lies above this line
the improvement is significant at more than the 99\% level.
The vertical line indicates the cooling radius (i.e. the 
radius at which the cooling time equals the Hubble time)
as determined from deprojection analysis of ROSAT images 
by Peres et al. (1998) for A1795 and by Allen (2000) for
A1835. For M87 all bins are within the cooling
radius ($r_{c}=20$ arcmin, Peres et al. 1998).

In the case of A1835 (Figure 1), the outermost radial bin that we consider
is completely outside the cooling region, thus the fact that 
neither model B nor model C provide a significant improvement with 
respect to model A is not at all surprising, as there is a general
consensus that the gas outside the cooling region is single-phase.
Indeed we have included in our analysis of A1835 and A1795 a radial bin 
outside the cooling region to use it as a sort of control point 
to verify that our statistical test is capable of confirming that
the gas is single-phase outside the cooling radius.
The middle bin in Figure 1 is partially in the cooling region and partially
outside so the fact that the improvement is not significant 
is not particularly compelling. In the innermost 
bin we find that both models B and C give substantially smaller $\chi^2$ 
than model A. In the case of model B the improvement is significant 
at slightly less  than the 99\% level while for model C it is 
significant at much more than the 99\% level. The rather large 
$\Delta \chi^2$ between model B and model C is telling us that the 
spectrum for this region is better fitted by an emission model 
characterized by a minimum temperature of  2.9$\pm 0.7$ keV 
(nicely consistent with the one derived by Peterson et al. 2001 from 
the analysis of the RGS spectrum), than by a model characterized by a 
temperature distribution extending down to 0.1 keV where most of the 
softer emission, which is coming from the cooler phases, is hidden by 
an intrinsic absorber.

In the case of A1795 (Figure 2), as for A1835, the outermost bin is almost
completely outside the cooling region, the $\Delta \chi^2$ for both
model B and C with respect to model A is zero, meaning that  
the multi-phase component is rejected by the fitting procedure by pushing
its normalization  to zero. The bin with bounding radii 1 and 2 arcmin
is completely inside the cooling region and for both model B and C 
the improvement with respect to model A is negligible. This is an
important result as it is telling us that the spectrum from
a region within the cooling radius, where the temperature profile
is not very different from the temperature outside the cooling-flow, does 
not show any evidence of being produced by a multi-phase gas. The next 
bin shows a qualitatively 
similar result, the  $\Delta \chi^2$ for both models B and C are not 
statistically significant at the 99\% level. Finally in the innermost bin, 
where, because of projection effects, the spanned temperature range
is largest, both models B and C show a significant improvement with 
respect to model 
A, but as for the innermost bin of A1835 the fit for model C is
substantially better than the fit with model B. 
The minimum temperature derived from model 
C is 1.9$\pm 0.5$ keV, which is consistent with what has been found by 
Tamura et al. (2001) from the analysis of the RGS data. 
Note also how the 
$\Delta \chi^2$ values increase continuously with decreasing radius, 
as is expected if the gas is single-phase and the multi-phaseness 
comes from putting together distinct regions with different 
temperatures and from the fact that, because of projection effects,
the spanned temperature range is larger at smaller radii.

M87 (Figure 3) is the nearest system we have investigated, indeed it is 
so near that the whole field of view is contained within the cooling 
radius. The temperature profile for this object (see B\"ohringer et al.
2001 and Figure 3) shows a small gradient for radii larger than $\sim $ 
2 arcmin and a rapid decrease for smaller radii.
The comparison of  $\Delta \chi^2$ in this cluster, which in some sense
is the most important, as it is the one where we can best separate
different emitting regions, points very clearly in favor of the
single-phase hypothesis. 
For the outermost bin (bounding radii 7-9 arcmin)
neither model B nor model C provides a substantially better fit 
than model A, indeed in the case of model B the multi-phase 
component is rejected by the fitting procedure itself
by pushing its normalization to zero.
Note that  this regions is  well within the cooling radius of M87 
($\sim$ 20 arcmin).
For all spectra at radii larger than 2 arcmin and smaller than 
7 arcmin,
model B does not give a better fit than the single-phase model, while
model C does. From the bottom panel of figure 3 it is clear
that the temperature range spanned by model C is always narrow,
thus the comparison between models A, B and C 
is telling us that a model containing a multi-phase component characterized 
by a narrow temperature range gives a better description of the 
data than a single temperature model, while a model containing a 
multi-phase component characterized by a broad temperature distribution
does not.  
At radii smaller than 2 arcmin, where the temperature profile 
suddenly steepens the $\Delta \chi^2$ shoot up. In the innermost two bins 
the implied statistical significance for both models is well over the 99\% 
level and, as for the innermost bin of A1795 and A1835, 
model C gives a much better fit than model B.

To better understand why model C performs so much better than model 
B we have made a detailed comparison between the best fitting 
model B and C spectra for  the innermost region of M87.
requiring temperatures 
In Figure 4 we show the observed spectrum with the best fitting multi-phase 
components for models B and C convolved with the EPIC instrumental response. 
The two components are  similar above 2 keV,
where the effect of the intrinsic absorption (model B), or of a 
cutoff temperature (model C) is modest. 
At smaller energies the difference becomes more evident.
In the region around 0.9 keV, where the Fe-L complex is dominant, the 
two spectra are very different, the model B component features a peak  
at 1 keV followed by  a broad shoulder extending down to about 0.7 keV, 
this is due primarily to Fe-L lines from low ionization states requiring 
temperatures smaller than about 1 keV; the model C component, 
as the model B component, presents a peak around 1 keV, but the decrease 
towards smaller energies is much more rapid because the low ionization
lines produced by gas at temperatures smaller than 1 keV are absent.
The $\Delta \chi^2$ between the best fits with model B and C over the entire 
spectral range considered is 113, with most of it,  $\Delta \chi^2 = 90$, coming 
from the energy range 0.7-1.2 keV, where the multi-phase components differ 
as explained above. 
In Figure 5 we show the residuals of the best fits with model
B and C, in this energy range. As can be seen model C is capable of 
reproducing the observed spectrum rather well, while model B is clearly 
in defect of the data between 0.9 and 1.0 keV and in excess between 0.7 and 0.9 keV, 
indicating that the multi-phase component in model B is characterized by 
an Fe-L shell quasi-continuum emission that it is too broad to correctly 
reproduce the data.
More specifically, there is too much emission from low ionization lines 
predominantly emitted from gas with temperature smaller than $\sim$ 1 keV 
and too little emission from higher ionization lines mostly coming from gas 
with temperatures larger than $\sim$ 1 keV. 
Thus the key feature allowing us to discriminate between models B 
(dashed line) and C (solid line), in the case of M87, is the shape of 
the Fe-L shell blend. 

The results of our spectral model comparison may be summarized 
as follows: for the outer regions, which are characterized by a narrow
temperature distribution, 
neither model B nor model C provide an improvement 
with respect to model A ( with the exception of the regions between 2 
and 7 arcmin in M87 where, due to the high statistical quality of 
the data, the fitting procedure can discriminate between a single 
temperature spectrum and a multi temperature spectrum characterized 
by a narrow temperature range); for the inner regions, which are 
characterized by a broad temperature distribution, models B and  C 
provide an improvement with respect to model A and model C always gives 
better fits than model B. 
These results all point in the same direction, namely that
the gas in cooling-flows does not show the broad temperature 
distribution expected from standard multi-phase models. The presence 
of a moderate temperature range, such as the one measured in 
the M87 spectra (see Fig. 3) could result either from a moderately
multi-phase gas or alternatively from a single phase gas 
characterized by an azimuthal temperature gradient.

\subsection{The intrinsic $N_H$}

In Figure 6 we plot the intrinsic $N_H$ profiles 
for A1835 (top panel), A1795 (middle panel) and M87 (bottom panel)
as derived from model B. 
If the physical scenario underlying model B were correct the  
intrinsic $N_H$ distribution should be characterized by a 
clear decrease with increasing radius for any realistic 
distribution of the intrinsic absorber, as the innermost 
emitting regions are viewed through all other regions and 
should therefore present the largest absorption.
The intrinsic $N_H$ distributions for all our clusters show 
no such trend.  The case of M87, where we have 7 bins and the errors 
are small, is particularly illuminating. 

If we accept that model B does not provide a realistic description 
of the data, then the  
large absorption column depths inferred by previous 
analysis of cooling-flow spectra (Allen 2000 and references 
therein) and from fits with model B presented in this paper,
 must be considered as an artifact.
Indeed if the gas is not strongly multi-phase, spectra 
for regions where the spanned temperature range  
is large can be fitted with a multi-phase model
only if the soft emission predicted by the standard multi-phase model, 
which is not present in the observed spectra, is somehow hidden. This 
can of-course be achieved by introducing an absorption component.
The good combination of spectral resolution, large spectral 
band and good spatial resolution of the EPIC cameras is 
allowing us, for the first time, to clearly discriminate between 
an absorbed, truly multi-phase spectrum (model B), and a "fake"
multi-phase spectrum characterized by a 
temperature range (model C).

We remark that model C, while certainly doing 
a better job than model B, does not provide
the most appropriate description of the spectrum. 
A model in which the projection effects, which contribute
to the multi-phase appearance of the
gas are explicitly accounted for would certainly give a 
better description allowing the study of the 
temperature structure of the ICM. We defer the analysis of our data 
with such a model to future work.

\subsection {Mass Deposition Rates}

Our analysis shows that 
 in  the outer rings  of all 3 clusters  the spectrum is 
consistent with being single-phase,
whereas inside 1 arcmin, 2 arcmin and 6 arcmin for A1835, A1795 and 
M87 respectively, model C provides a substantially better fit than 
model A. Model C also provides a better fit than model B, implying
that the spectral shape in the soft band is always better
modeled by a minimum temperature than by an intrinsic absorber. 
The minimum temperature is always larger than about 1 keV and, 
in the case of A1835 and A1795, it is consistent with that derived from 
the analysis of RGS data.
Thus, for all cooling-flow cluster spectra investigated with XMM-Newton 
so far, either with EPIC or RGS, we find compelling
evidence for a lack of emission from gas with temperatures
smaller than 1-2 keV. 
If we interpret this result in the most simple and straightforward way, 
that is by assuming that gas does not cool below
1-2 keV, then some profound implications follow.
Indeed, if the gas does not cool below a given temperature it
will not be deposited in the form of cold gas.
Thus any mass deposition probably occurs on scales smaller than the ones 
we have investigated and the mass deposition rates of 
cooling-flow clusters may be much smaller than previously thought.
Under the above assumptions, the $\dot M$ derived with model B from 
the analysis in the innermost bin for all our clusters can be regarded 
as an upper limit to the total mass deposition rate, as we expect
that most of the multi-phase component is actually due to the 
mixing of gas from physically distinct emission regions.
In the case of A1835 we find $\dot M = 247.8\pm 133 M_\odot$/yr
which is to be compared with $1150\pm 450 M_\odot$/yr 
from the deprojection of ROSAT PSPC images (Allen 2000);
for  A1795 we find $\dot M = 27.6 \pm  6.2  M_\odot$/yr
against $450\pm 50 M_\odot$/yr (Peres et al. 1998),
and in the case of M87 $\dot M < 0.5  M_\odot$/yr
to be compared with $40\pm 5  M_\odot$/yr (Peres et al. 1998).

\section {Discussion}

In a recent paper Fabian et al. (2001) proposed three
possible ways of reconciling 
the absence of emission lines in the RGS spectra of 
cooling-flow clusters with the presence of gas with 
temperatures smaller than 1-2 keV. 
In the following we discuss each of these possibilities
in the light of our results on the EPIC data.

The first solution outlined by Fabian et al. (2001) 
is to hide the emission lines behind a patchy intrinsic absorber, 
possibly more concentrated towards the center of the cluster.
From Figure 6 and the related discussion we have evidence that 
if such an absorber exists it probably operates on scales smaller than 
the ones we can investigate with EPIC. If that is the case then 
mass deposition will occur only in the innermost regions
and not throughout the entire cooling-flow region.
 
Another possibility considered in Fabian et al. (2001)
would be for the gas to be highly bimodal
in its metal distribution with say 10\% of the gas having
a metallicity of 3 in solar units, and the remaining
90\% having no metals at all. At temperatures below
about 2 keV, when line cooling becomes important,
the metal rich gas would cool at a much faster rate 
than the metal poor gas and lines from metals
at temperature below about 2 keV would not be observed.
We have tested this possibility by allowing 
the metallicity of the multi-phase component in model
B to vary independently from the metallicity of 
the single-phase component.  
In none of our fits do we find the best fitting value of the
metallicity of the multi-phase component to be significantly 
smaller than the one of the single-phase component, implying 
that the multi-phase gas is not metal poorer than the 
ambient gas. If this mechanism is operating,
it must be doing so on scales smaller than the ones 
we have investigated and mass
deposition should be confined to such scales.  

The third possibility indicated by Fabian et al. (2001)
is that the X-ray emitting gas may be cooling very rapidly
by mixing with gas at $10^3$ K, which is known to 
be present within about 50-70 kpc of the center of some cooling-flow
clusters. However, as pointed out by Crawford et al. (1999),
line emitting gas at about $10^3$ K is not always found in the 
core of cooling-flow clusters, and in most clusters 
it extends to radii smaller than the cooling radius. In the
case of A1795, the filaments extend out to about 20 kpc from 
the core (Hu et al. 1985), while the cooling radius is 
about ten times larger (Peres et al. 1998). 
Thus, at least in the case of A1795, it is 
unlikely that this mechanism could allow mass deposition 
to occur at radii larger than  20 kpc where, according 
to multi-phase models, most of the mass deposition should 
be taking place.
Moreover, if this mixing of hot with cold gas mechanism 
does operate, 
it will most likely end up depositing cold gas in the 
cooling flow region.
The physical scenario is one in which the cooling flow 
component should be characterized by a minimum temperature 
and should also present substantial absorption from the 
gas which has cooled from X-ray emitting temperatures.
We have tested this possibility
by introducing an intrinsic absorber acting on the 
cooling flow component in model C.  By performing fits in
the regions where model C gives substantially better fits 
than model A we find 
that the best fitting values of the intrinsic absorber 
are always small, typically a few times $10^{20}$ cm$^{-2}$.
In M87 where the statistics is best, for most regions the
intrinsic absorber has a column depth smaller than 
$10^{20}$ cm$^{-2}$.  
Such small values are inconsistent with the deposition of large 
amounts of cold gas over the putative lifetime of the cooling-flow. 

If we accept that mass deposition in cooling-flow clusters
is smaller and confined to smaller scales 
than previously thought then the long standing 
problem of the lack of evidence for large amounts of cold gas 
in cooling-flows, from observations carried out in bands other 
than X-rays, is solved.
Investigations longwards of X-rays never found substantial
amounts of gas because the gas was never there in the first 
place.   

If there is  no mass deposition 
at large radii it would seem reasonable to assume that all 
the cooling gas must flow to the center. However the surface 
brightness profile for such a cooling-flow cluster would be 
much more centrally peaked  than is commonly observed 
(see Fabian 1994 and refs. therein).
Indeed multi-phase models have been originally developed 
(Nulsen 1986) to explain why the surface
brightness profile is not as peaked as would be expected
if all the gas flowed to the center of the cluster.
Perhaps the most obvious way out of this dilemma is 
to assume that a heating mechanism is 
countering the radiative cooling. Over the past 15 years
many different heating mechanism have been proposed 
(e.g. Tucker \& Rosner 1983, Binney \& Tabor 1995, Ciotti \& Ostriker 2001).
However no stable heating process yet devised is able
to counteract the effects of radiative cooling and account for
observed X-ray images (Canizares et al. 1993). This has in turn led 
some workers to consider sporadic heating mechanisms. Several heating 
mechanisms of this kind have been presented in the past.
Recently Soker et al. (2001), motivated by the analysis 
of Hydra A with Chandra data (David et al. 2000), 
have proposed heating by sporadic outbursts of the central radio 
source. The fact that A1835 and Sersic 159-03 do not have strong
radio sources, in our view, argues against such a mechanism.
Indeed we should look for a heating mechanism  
that can operate in all cooling clusters.


Before XMM-Newton and Chandra the lack of a 
stable heating process that could counteract the effects 
of radiative cooling and account for observed X-ray images
was considered as a crucial point in favor of multi-phase
models. 
Now that the new results emerging from the analysis of XMM-Newton and 
Chandra data question the standard cooling-flow picture,
the hunt for a self consistent heating mechanism is again open.

Last, but perhaps not least, the term "cooling-flow"
cluster is most likely no longer appropriate, as it describes 
a phenomenon of smaller entity and impact than previously thought. 
We propose to substitute it with that of "cool-core" cluster. The latter
definition is less ambitious than the first, as it reflects only an 
observational fact rather than an inferred physical property, the flow, 
but has the undeniable advantage of being firmer. 


\acknowledgments
We thank the many people who have contributed to building, 
calibrating and operating the EPIC instrument on-board XMM-Newton.
We thank S. De Grandi and  S.Ghizzardi for a critical reading of 
the manuscript.



\clearpage


\begin{figure*}[htb]
\centering
\includegraphics[angle=-90,width=13.5cm]{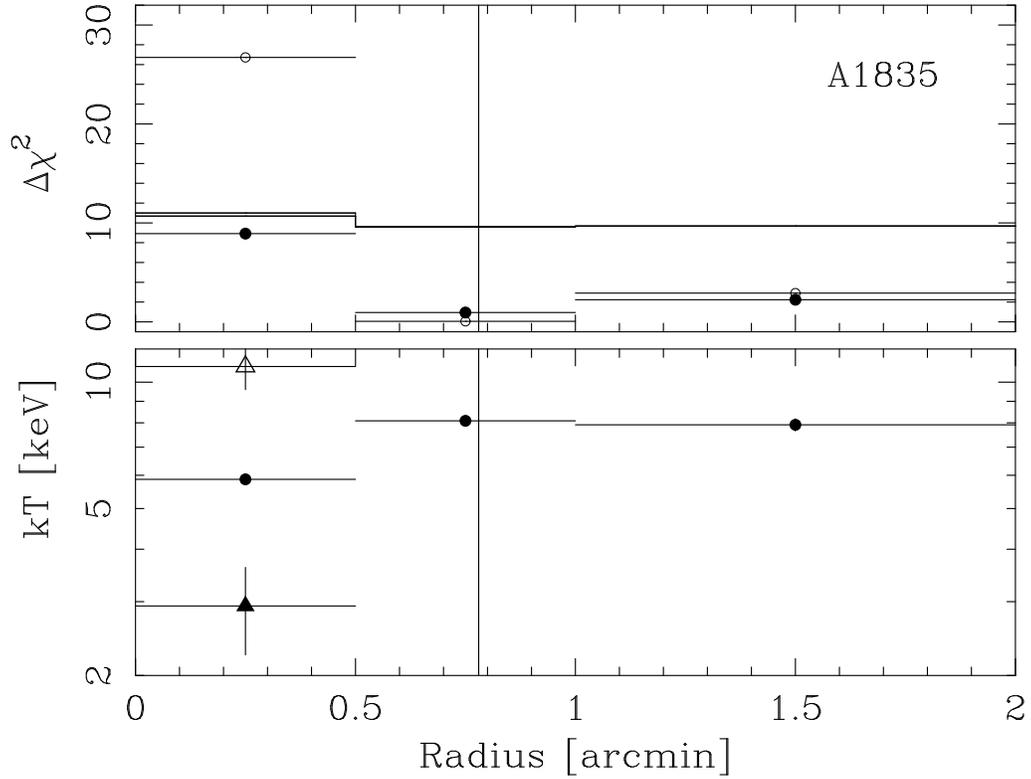}
\caption{
{\bf Top Panel}: 
$\Delta \chi^2$ between models A and B (filled circles) 
and models A and C (open circles) as a function of radius.  
The horizontal lines indicate 
the $\Delta \chi^2$ values for which the statistical improvement of 
the model B (C) fit with respect to model A are significant at the 
99\% level according to the F-test.
The vertical line indicates the cooling radius (i.e. the 
radius at which the cooling time equals the Hubble time).
{\bf Bottom Panel}:  Temperature profile for 
A1837 as obtained from model A, filled circles . 
For bins where model C gives a better fit than model A 
we also show the maximum and minimum temperatures from model C, 
which are indicated as empty and full triangles
respectively. 
Uncertainties are at the
68\% level for one interesting parameter
($\Delta \chi^2 = 1$).
}
\end{figure*}


\begin{figure*}[htb]
\centering
\includegraphics[angle=-90,width=13.0cm]{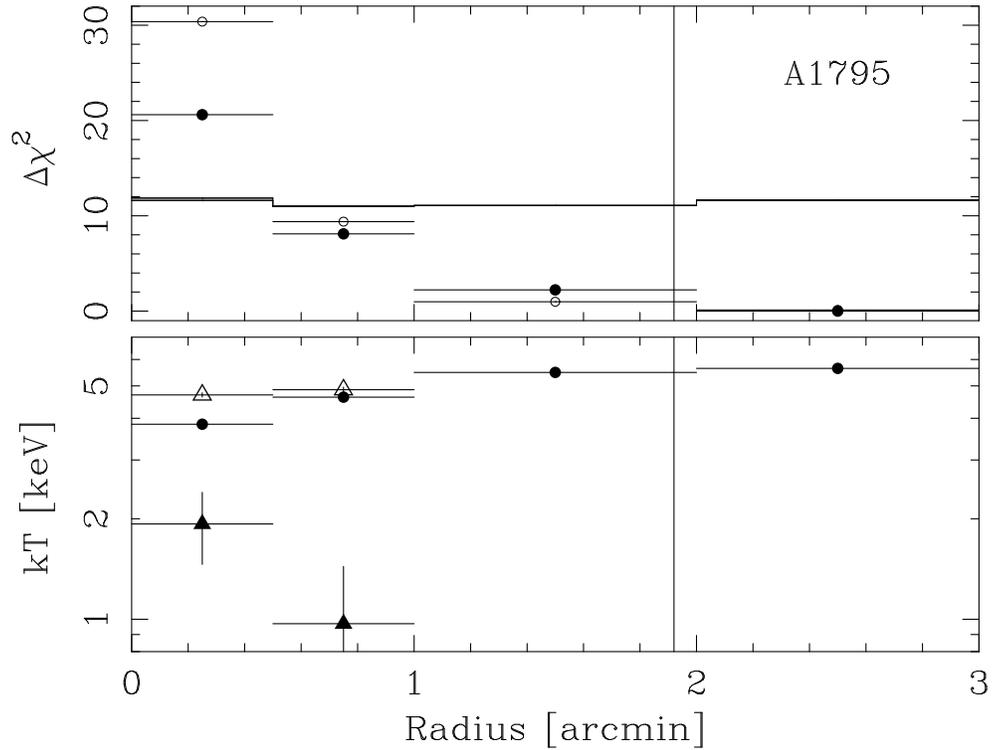}
\caption{
Same as Fig. 1, for A1795.
}
\end{figure*}


\begin{figure*}[htb]
\centering
\includegraphics[angle=-90,width=13.0cm]{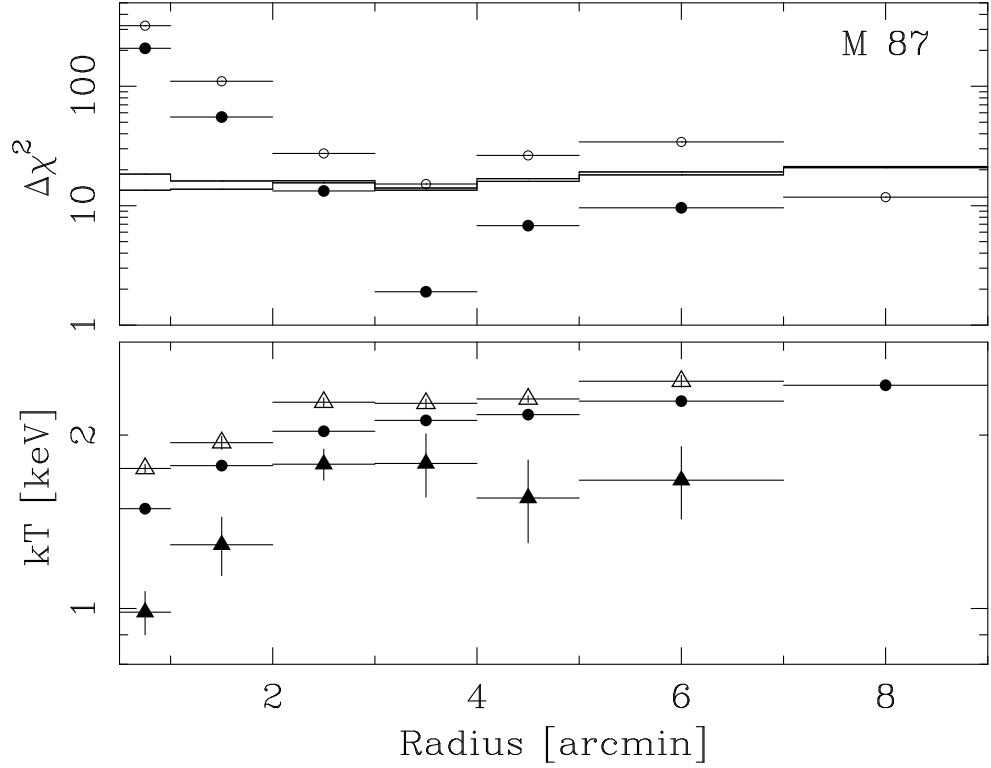}
\caption{
Same as Fig. 1, for M87.
Note that the $\Delta \chi^2$ value obtained by comparing model A to model B
in the outermost bin, which  is 0, is not shown in the plot.
} 
\end{figure*}


\begin{figure*}[htb]
\centering
\includegraphics[angle=-90,width=13.0cm]{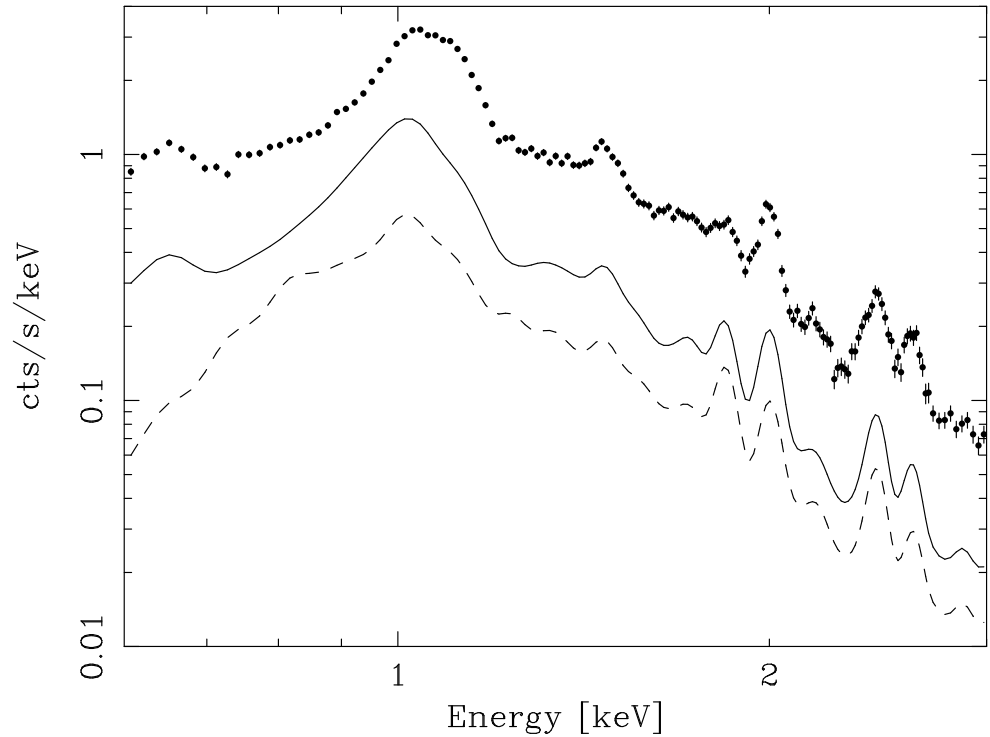}
\caption{
M87 spectrum for the annulus with bounding radii 0.5-1.0 arcmin.
The filled circles are the datapoints, the dashed and solid lines
represent respectively  the best fitting
multi-phase components for models B and C convolved with the EPIC 
instrumental response.
}
\end{figure*}


\begin{figure*}[htb]
\centering
\includegraphics[angle=-90,width=13.0cm]{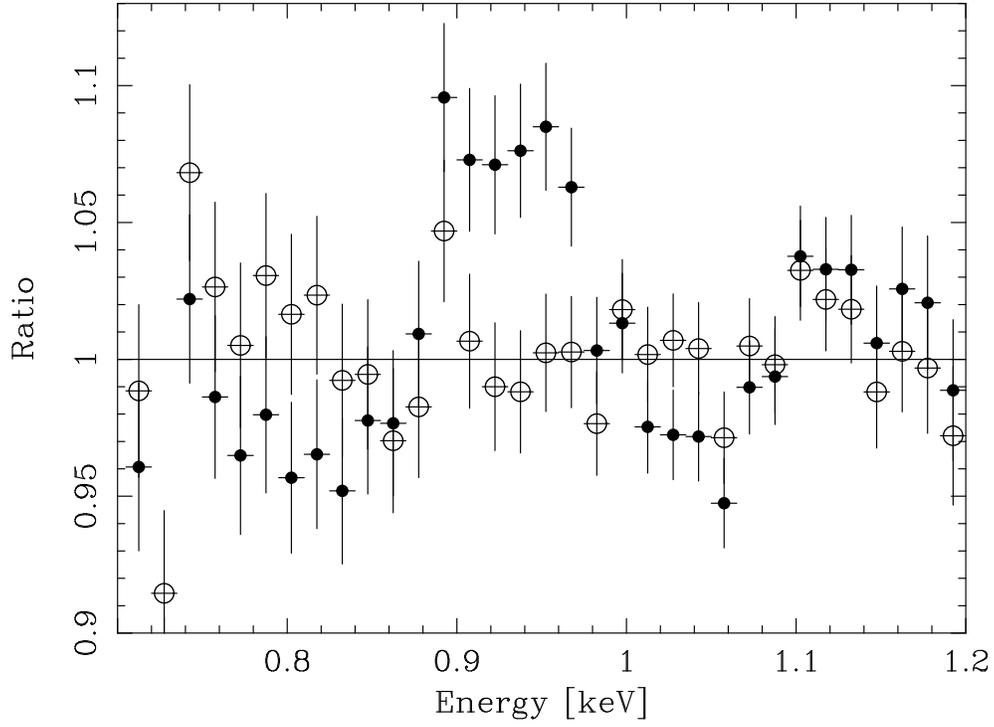}
\caption{
Residuals in the form of a ratio of the data over the model for 
the M87 spectrum shown in Figure 5.
The filled and open circles are respectively the residuals for models B
and C.
}
\end{figure*}


\begin{figure*}[htb]
\centering
\includegraphics[angle=-90,width=13.0cm]{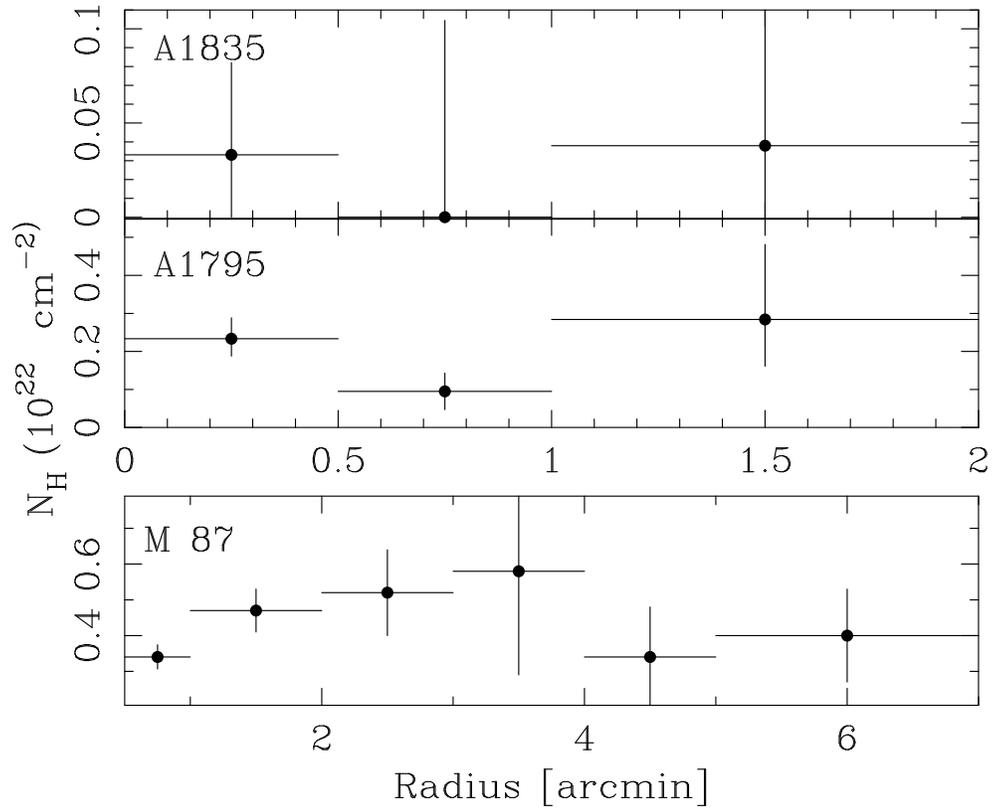}
\caption{
Radial intrinsic absorption profiles as obtained with model B, for A1835 
(top panel), A1795 (middle panel) and M87 (bottom panel). Spectral fits 
beyond 2 arcmin for A1835 and A1795 all have $N_H$ confidence 
intervals ranging from 0 to at least 10$^{22}$ cm$^{-2}$.
}
\end{figure*}

\end{document}